\documentclass[runningheads]{llncs}
\usepackage[T1]{fontenc}

\usepackage{comment}
\usepackage{xurl}
\usepackage{xcolor}

\usepackage{listings}
\usepackage[T1]{fontenc}
\usepackage{multirow}
\usepackage[inkscapelatex=false]{svg}

\begin{document}
\title{CI/CD Efforts for Validation, Verification and Benchmarking OpenMP Implementations}
%
%
\author{Aaron Jarmusch\inst{1}
\and
Felipe Cabarcas\inst{1,3}
\and
Swaroop Pophale\inst{5}
\and
Andrew Kallai\inst{1}
\and 
Johannes Doerfert\inst{2}
\and 
Luke Peyralans\inst{4}
\and
Seyong Lee\inst{5}
\and
Joel Denny\inst{5}
\and
Sunita Chandrasekaran\inst{1,6}
}
\authorrunning{Jarmusch A. et al.}

\institute{University of Delaware, Newark, DE, USA
\and
Lawrence Livermore National Laboratory, Livermore, CA, USA
\and
Universidad de Antioquia, Medellin, Colombia
\and
University of Oregon, Eugene, OR, USA
\and
Oak Ridge National Laboratory, Bethel Valley Road
Oak Ridge, TN, USA
\and 
\email{\{schandra\}@udel.edu}
}

\maketitle              
\begin{abstract}
Software developers must adapt to keep up with the changing capabilities of platforms so that they can utilize the power of High-Performance Computers (HPC), including exascale systems. OpenMP, a directive-based parallel programming model, allows developers to include directives to existing C, C++, or Fortran code to allow node level parallelism without compromising performance. This paper describes our CI/CD efforts to provide easy evaluation of the support of OpenMP across different compilers using existing testsuites and benchmark suites on HPC platforms. Our main contributions include (1) the set of a Continuous Integration (CI) and Continuous Development (CD) workflow that captures bugs and provides faster feedback to compiler developers, (2) an evaluation of OpenMP (offloading) implementations supported by AMD, HPE, GNU, LLVM, and Intel, and (3) evaluation of the quality of compilers across different heterogeneous HPC platforms.
With the comprehensive testing through the CI/CD workflow, we aim to provide a comprehensive understanding of the current state of OpenMP (offloading) support in different compilers and heterogeneous platforms consisting of CPUs and GPUs from NVIDIA, AMD, and Intel.
\keywords{OpenMP \and Validation \and CI \and Compiler.\and Benchmarking}
\end{abstract}
%
%
%
\section{Introduction}
\label{sec:introduction}
Heterogeneous computing has reached a new milestone with the Frontier \cite{Frontier} and Aurora \cite{Aurora} supercomputers achieving exaflops, representing one quintillion floating point operations per second.
 This significant increase in processing speed has the potential to revolutionize application performance, provided that software developers can keep up with the growing demand for tools and platforms that can harness the power of these devices. To achieve this, current languages and models include  OpenMP \cite{openmp52}, OpenACC \cite{OPENACC}, CUDA \cite{CUDA}, HIP \cite{bauman2019introduction}, and OpenCL \cite{OPENCL}. However, each of them has its own unique features and their usage depend on what developers are comfortable with. 

While some languages like NVIDIA's CUDA, and AMD's HIP, can be challenging for beginners to learn and require significant rewriting of programs to achieve optimal performance, portable programming models such as OpenMP and OpenACC offer a simpler approach. They use a directive-based approach for parallel programming, allowing users to include these directives on top of existing C, C++, or Fortran code, without compromising on performance. These models play a major role with heterogeneous systems equipped with  accelerators such as GPUs, FPGAs, APUs, and more, and can be parallelized on ARM-based systems with ease. 

Since more real-world applications have adopted OpenMP, it has been used in various domains. For instance, miniQMC~\cite{huber2022efficient}, primarily used in the study of electronic molecular structures and 2D/3D solid states, is a simplified version of QMCPACK~\cite{kim2018qmcpack}. Other applications include miniMD~\cite{pennycook2018evaluating}, LULESH~\cite{karlin2016early}, GAMESS~\cite{bak2022openmp}, HPGMG~\cite{daley2020case}.
With the adoption of OpenMP in modern computing, it is crucial to validate and verify the compilers' implementations of this standard. 
To simplify the evaluation process for developers, we have established a Continuous Integration (CI) and Continuous Development (CD) workflow, which provides faster feedback for developers. 
This approach allows us to evaluate the support of OpenMP across different compilers more frequently, reducing the waiting time for developers between evaluations. By implementing this workflow, our goal is to improve the development process and ultimately make OpenMP easier for developers to use in their applications. 

This work's focus is on the CI/CD efforts to evaluate the support of OpenMP across different compilers using established validation suites and benchmarks such as OpenMP Validation and Verification (V\&V) suite \cite{OMPVV}, SPEChpc \cite{SPEChpc}, HeCBench \cite{HECBENCH}, and Smoke \cite{SMOKE} tests. 
The main contributions of this paper are:
\begin{itemize}
    \item Building a Continuous Integration and Continuous Development pipeline for automating testing of OpenMP implementations
    \item Analysis of output from the CI/CD pipeline that includes verification of OpenMP implementations from AMD, HPE, GNU, LLVM, NVIDIA, and Intel
    \item Discussion and evaluation of the performance and quality of OpenMP offloading compiler implementations on supercomputers such as Frontier, Perlmutter, Sunspot, and Summit using SPEChpc benchmarking suite.
     
\end{itemize}


\section{Background \& Motivation}
\label{sec:background}
OpenMP is a widely used application programming interface (API) that enables programmers to develop parallel applications in C, C++, and Fortran with access to multi-platform shared memory and multi-processing capabilities. With its straightforward adaptable interface, OpenMP enables developers to create efficient parallel programs on multi-core processors. Key features of OpenMP include directives for parallel programming, runtime library routines, thread management, and environment variables. 

Over the years OpenMP has refined support for computations on CPU and, starting with OpenMP 4.0 \cite{openmp40}, added support for devices such as accelerators. Although GPGPUs are most widely used accelerator devices used in HPC, OpenMP directives are designed to work with any devices that have memory and are capable of performing computations. That makes OpenMP an attractive choice for future-proofing codes and minimizing developer effort for adapting to different architectural trends.

With multiple vendor implementations of OpenMP and more and more real-world applications porting/developing codes using OpenMP programming model, it is crucial to validate and verify the compilers' implementations of the OpenMP standard. It is often unclear to the application developers if adequate OpenMP features are supported by an implementation and if they are performant. This effort takes that stress away from developers by providing nightly runs of established and curated set of benchmarks and testsuites.

\section{Related Work}
\label{sec:related}
The current work's focus is to enhance the testing of OpenMP offloading implementations using a CI/CD workflow, which has become increasingly popular in recent years due to its ability to automate various stages of the development process~\cite{ULexperience,10.1145/3631519}. 
Clacc~\cite{denny2018clacc}, an open source project that offers support of OpenACC in Clang and LLVM, employs a CI/CD workflow to detect errors quickly. However, the main focus of the current work is to bring a workflow to OpenMP implementation. Even though both OpenACC and OpenMP are directive-based models, their compiler implementations differ, resulting in distinct use cases. 
LLVM Buildbots\footnote{https://lab.llvm.org/buildbot/}, another well-known project in the LLVM community, has been used for commit checking of LLVM, but not exclusively for testing OpenMP. However, there are only a few buildbots dedicated to testing OpenMP Offloading, highlighting an area of opportunity for further development. 
Additionally, CI/CD workflows have also been applied to Machine Learning (ML) through MLOps. Toward MLOps~\cite{9582569}, is a case study demonstrating the use of CI/CD workflows to train ML models with different configurations, analyzing the results to identify potential performance bottlenecks and improve the process. 
Finally, CI/CD workflows have been utilized on the RMACC Summit supercomputer~\cite{rmacc} for deploying user-facing software environments and automating benchmark testing~\cite{10.1145/3219104.3219147}.
Overall, the current work builds upon these existing efforts to create a more comprehensive and efficient CI/CD workflow for OpenMP Offloading implementations.

\section{Continuous Integration and Continuous Development Workflow}
\label{sec:CICD}
Manual software testing methods are insufficient for hardware evolution and new software releases. To address this challenge, we use an automation tool to test compilers when they are updated. Leveraging a CI/CD workflow allows us to continuously identify failed tests at all times, not just before an official software release, and to automate the manual testing process. As a result, we reduce the time and effort required for compiler testing, leading to faster development cycles and higher quality software.

\subsection{CI/CD Pipeline}

We have established an OpenMP CI/CD workflow that involves setup of source code, building and installing the compiler, and suite execution, to reduce the developers time required for testing. 

Within our CI/CD workflow a \textbf{pipeline} refers to a single run of our workflow, which can be triggered by a scheduled time or by new commits. A pipeline contains \textbf{stages} which defines when to run a job. A \textbf{job} is a set of steps that defines what you want to accomplish. 
A pipeline's success depends on the completion of each job, which are labeled as "Pass" or "Fail." If any job within a stage fails, the entire pipeline fails. In other words, each job serves as a building block for the overall pipeline, and its outcome determines whether the pipeline is successful or not. 
The separation of jobs within a pipeline allows isolation of issues from different parts of an implementation during the collection of source code (setup), build/install of the compiler, or suite execution. 

\begin{figure}[h]
    \centering
    \includesvg[width=0.9\textwidth]{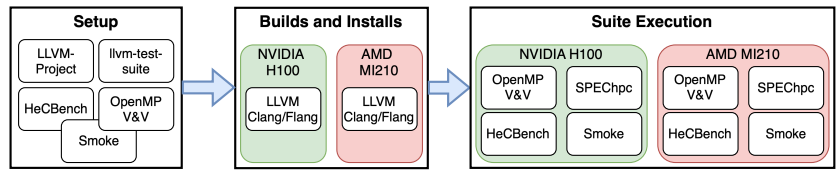}
    \caption{CI/CD Workflow for LLVM Clang and Flang}
    \label{fig:stages}
\end{figure}

With this idea of separation, as indicated in Figure~\ref{fig:stages}, our workflow (for LLVM Clang/Flang) is divided into three stages: setup, build, testing followed by cleanup. 

\textbf{In the first stage (i.e., setup), }the CI/CD clones any necessary source codes such as the LLVM source, LLVM testsuite infrastructure, and other suites (OpenMP V\&V, Smoke, and HeCBench). However, we do not clone SPEChpc suite because the benchmark is not released via GitHub or any other public software storage platform. 
{https://www.spec.org/hpgdownload.html}.
We download the source codes at a specific commit (for OpenMP V\&V, Smoke and HeCBench) to forgo any new changes coming into the suite after the specific commit we have chosen. For the LLVM compiler source, we download the commit (GitHub commit hash) from the latest trunk. 

\textbf{In the build/install stage,} that comes next, we build only the LLVM compiler from the identified commit. We then, proceed to testing with OpenMP V\&V, Smoke, HeCBench as well as SPEChpc suite.  

\textbf{In the suite execution stage,} the CI/CD is using the LLVM Integrated Tester (LIT)~\cite{LIT} in conjunction with the LLVM-test-suite~\cite{LLVM-TestSuite} to compile and execute the OpenMP V\&V, HeCBench, and Smoke suites. We designed CMake files for each testsuite, located in the LLVM-test-suite GitHub repository\footnote{https://github.com/llvm/llvm-test-suite/tree/main/External}. These CMake files allow us to specify which languages we want to run and which tests or applications to execute. For instance, the OpenMP V\&V testsuite consists of a large number of tests, and to make a job pass or fail, we need to provide specific criteria for the CI/CD to determine this. Therefore, we manually ran each suite to create a green and red list of tests for capturing pass and fail results respectively, with each compiler. Additionally, we created green and red lists per accelerator that we target.

In our CI/CD pipelines, we only run the green lists of tests, which ensures reliable and accurate testing results. By integrating both correctness checking and real-world application testing into a single workflow, we can ensure that our compilers are not only correct but also reliable in real-world scenarios. 

\textbf{Experimental Setup:} Since hardware evolves at a rapid pace, The University of Oregon (UO) has setup the Frank Cluster which is comprised of a multitude of servers hosted by UO Oregon Advanced Computing Institute for Science and Society (OACISS) with different hardware setups. At the time of writing, across different servers they have an NVIDIA H100, AMD MI210, and an Intel Data Center Max 1100 (Ponte Vecchio). Utilizing the Frank Cluster to implement our first CI/CD to test software means we can easily change to new hardware configurations. 
For our CI/CD, we are using GitLab and focusing on the AMD MI210 and the NVIDIA H100 within the Frank Cluster. This means we build compilers and execute each suite for both AMD and NVIDIA GPUs for target-specific errors. For instance, we are building LLVM Clang and Flang for AMD and NVIDIA GPUs, which makes two jobs. We also run each suite on both AMD and NVIDIA GPUs. In total, this makes eight jobs. Two and six for the build and testing stages respectively. 

\begin{figure}[h]
\vspace{1mm}
    \centering
    \includegraphics[width=0.9\textwidth]{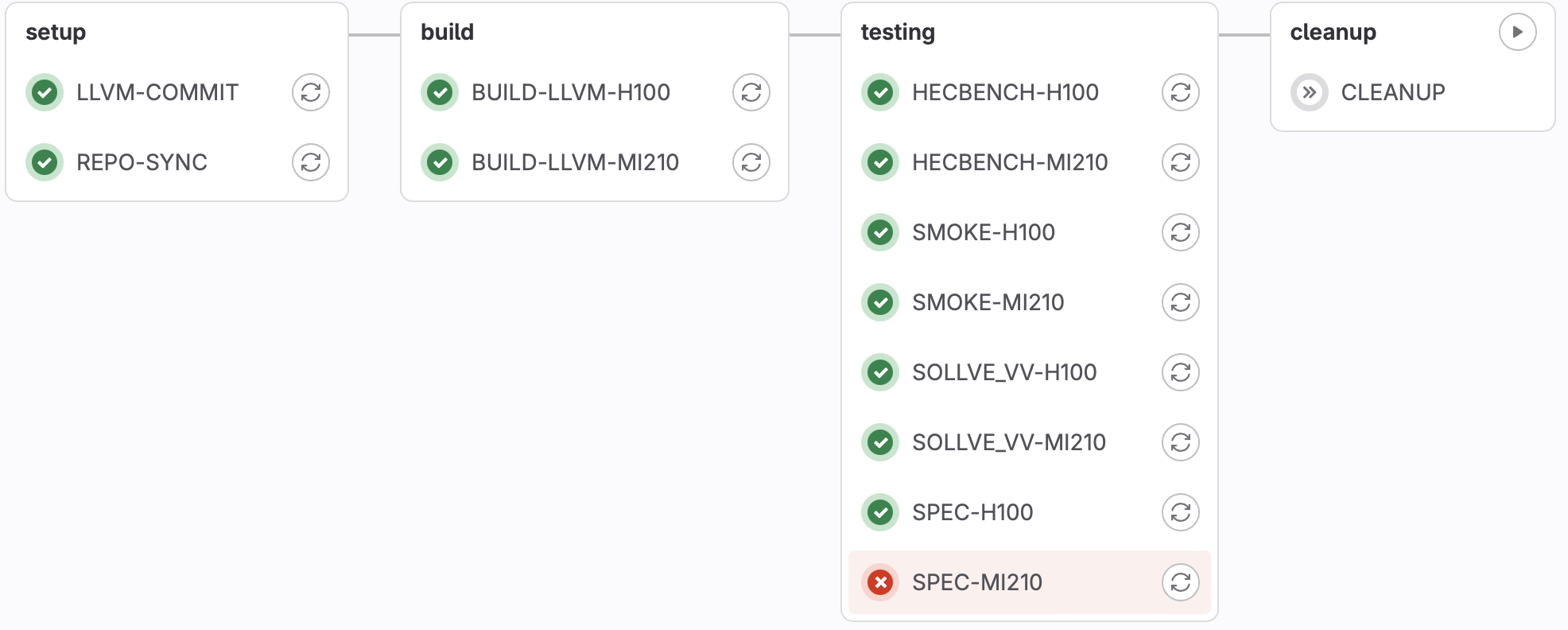}
    \caption{The CI/CD output for a LLVM Clang and Flang pipeline. The green check-mark marks the job as "Pass" while the red 'x' marked the job as "Fail".}
    \label{fig:CI-output}
    \vspace{-5mm}
\end{figure}

On the Frank Cluster we test LLVM Clang and Flang hourly with the OpenMP V\&V, HeCBench, and Smoke suites. We run the SPEChpc suite weekly, due to long execution times. The hourly testing is done to provide commit-level feedback to developers because LLVM is open-source and most compiler developers upstream LLVM into their own compiler~\cite{10024603}. All other compilers are tested on a weekly basis with the SPEChpc suite, since the frequency of compiler releases or changes to those compilers are more on a monthly basis. Table~\ref{tab:CIconfiguration} is our configuration with the compilers being tested on what hardware. Since we wanted to include testing for Cray we reproduced our CI/CD workflow onto Frontier at ORNL. This also gave us another system to validate our results against. An example of the CI/CD pipeline is shown in Fig.~\ref{fig:CI-output}. 
\begin{table}[]
    \caption{CI/CD pipeline hardware and software configuration.} 
        \label{tab:CIconfiguration}
    \centering
\begin{tabular}{|l|l|c|c|c|c|}
\hline
\multicolumn{2}{|c|}{\textbf{System}} & \multicolumn{4}{c|}{\textbf{Compilers}}    \\ \hline \hline
UO Frank Cluster & H100 & LLVM Clang \& Flang & GNU & NVIDIA & - \\ \hline
UO Frank Cluster & MI210 & LLVM Clang \& Flang & GNU & AMD & - \\ \hline
ORNL Frontier    & MI250X & LLVM Clang \& Flang  & GNU & AMD & Cray\\ \hline
\end{tabular}
\end{table}

\textbf{Discussion:} 
The development and testing of compilers require rigorous testing to ensure quality. While automating these tasks can accelerate the process, it's crucial to maintain effective communication with developers to provide timely feedback on testsuites' failures.
To address this challenge, we have implemented two forms of automatic communication. First, for LLVM, we have integrated a messaging service to notify a team of compiler developers interested in LLVM's implementation of OpenMP offloading when a pipeline has failed. This provides the developers with immediate feedback in case a recent commit has broken the compiler. However, due to LLVM's open-source nature and large project aspect, it may not be possible to identify the root cause of a build failure solely through automated means, it may or may not be the fault of LLVM Clang or Flang OpenMP offloading implementation. The cause may have risen from other parts of the LLVM toolchain. Therefore, we still require some form of human interference to differentiate between failed jobs and identify the specific issue. 

Similarly, build-bots have been used in LLVM  to test every commit. These build-bots notify the commit creator if a failure occurs; however they primarily focus on CPU code testing and building. Consequently, our primary concern is not build failures but failures within the testsuites themselves. Therefore, it might not be advantageous to notify commit creators of these failures. Instead, we should track them specifically for developers working on OpenMP Offloading. 

For the second form of communication, we are utilizing the LLVM Nightly Testing (LNT)~\cite{LNT} infrastructure to create a server that stores testsuite information about the kernel. This information is visualized in a graph to track performance over time\footnote{https://crpl.cis.udel.edu/lnt-sollve/}. 
Overall, these methods of communication with developers about the results of the pipelines have been effective so far. However, we continue to evaluate and improve our approaches to ensure optimal effectiveness in the development and testing of compilers.

\section{OpenMP Validation \& Verification Suite and SPEChpc Benchmarking Suite}
In this section, we will elaborate further on the OpenMP V\&V testsuite and SPEChpc benchmarking suite exercised by our CI/CD workflow. The OpenMP V\&V testsuite is primarily concerned with verifying compiler correctness, whereas the SPEChpc benchmarking suite evaluates the quality of compiler implementations. By integrating both correctness verification and application-based benchmarking into a CI/CD workflow, we can ensure more robust and reliable compiler testing. 


\subsection{OpenMP Validation and Verification Suite}
\label{sec:VV}
The OpenMP Validation and Verification (OpenMP V\&V) 
suite~\cite{OMPVV,10024615} started as a sub-project of the SOLLVE (Scaling OpenMP With LLVM for Exascale) Exascale Computing Project (ECP)~\cite{sollve}. The sub-project was born out of the lack of open, unbiased testsuite to evaluate aspects of OpenMP that were most critical for exascale applications. 
Yet, there is still more work to be done. ECP marked a significant achievement with the push towards exascale computing resources. ECP ended in December 2023. As a path forward, the Next Generation Science Software (S4PST) project~\cite{s4pst} aims to push the boundaries even further. This focuses on creating a more predictive ecosystem for sustainability in HPC software, with a particular emphasis on node-level programming systems and tools. 
The tests in the OpenMP V\&V are organized based on the specification version. They contain feature and kernel tests in C, C++ and Fortran. 


\textbf{Coverage:} Table \ref{tab:numTests} shows there are 742 total tests in the testsuite covering new features starting with the OpenMP 4.5 specification up to the 5.2. The OpenMP testsuite is under active development and the numbers here represent a snapshot at the time of this writing.
The number of test per specification is mainly related to the number of new features introduced in the specification.

\begin{table}[]
    \caption{Number of tests in C, C++, and Fortran for each OpenMP Specification currently in the OpenMP Validation and Verification Suite.} 
    \centering
\begin{tabular}{|l||c|c|c|c|}
\hline
Version & C   & C++ & Fortran & Total \\ \hline \hline
4.5     & 134 & 14  & 104     & 252   \\ \hline
5.0     & 191 & 13  & 128     & 332   \\ \hline
5.1     & 99  & 2   & 28      & 129   \\ \hline
5.2     & 16  & 8   & 5       & 29    \\ \hline
Total   & 440 & 37  & 265     & 742   \\ \hline
\end{tabular}
    \label{tab:numTests}
\end{table}

\begin{figure}[h]
    \centering
    \includesvg[width=0.9\textwidth]{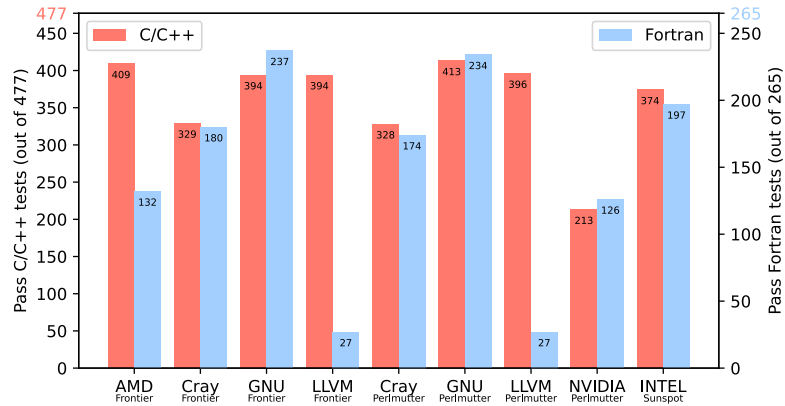}
    \caption{No. of tests passing per system out of total 477 C/C++ \& 265 Fortran tests}
    \label{fig:pass-tests}
        \vspace{-4mm}
\end{figure}

\begin{table}[]
\caption{Systems and compiler versions used for validation } 
    \centering
\begin{tabular}{|l|c|c|p{0.5\linewidth}|}
\hline
\textbf{System} & \textbf{Vendor} & \textbf{Accelerator} & \textbf{Compiler and Versions}  \\ \hline
Perlmutter     & HPE & NVIDIA A100  & NVIDIA 24.5, Cray 17.0.0, LLVM 19.0.0 commit (18ec885a), and GNU 14.1   \\ \hline
Frontier     & HPE & AMD MI250X  & AMD's ROCm 6.0.0, Cray 17.0.0, LLVM 19.0.0 commit (18ec885a), and GNU 14.1   \\ \hline
Sunspot     & Intel  & Intel GPU Max Series   &  OneAPI 18.0.0   \\ \hline
\end{tabular}
    \label{tab:systems}
    \vspace{-5mm}
\end{table}
\textbf{Results:} Figure \ref{fig:pass-tests} shows the number of tests that pass for each compiler in each system, for C/C++ and Fortran separately\footnote{\url{zenodo.org/doi/10.5281/zenodo.12571032}: allCompilerSystemsResults.json}. 
The left axis represents C/C++, while the right is Fortran. AMD C/C++ compiler on Frontier and GNU on Perlmutter pass the most, with GNU is the compiler that passes most tests over all. In the case of Fortran, in both systems GNU passes the most tests. LLVM Flang for offloading is under development. Table \ref{tab:perVersionResults} shows the results for each OpenMP Version.

\begin{table}[h]
\caption{OpenMP Validation and Verification Suite pass results per OpenMP Version.} 
    \centering
\begin{tabular}{|c|c|cccc|cccc|c|}
\hline
\multicolumn{1}{|l|}{} & \multicolumn{1}{l|}{}    & \multicolumn{4}{c|}{\textbf{Frontier}}                                                                                     & \multicolumn{4}{c|}{\textbf{Perlmutter}}                                                                                      & \textbf{Sunspot} \\ \hline
\textbf{Ver.}          & \textbf{Lang.}           & \multicolumn{1}{c|}{\textbf{AMD}} & \multicolumn{1}{c|}{\textbf{Cray}} & \multicolumn{1}{c|}{\textbf{GNU}} & \textbf{LLVM} & \multicolumn{1}{c|}{\textbf{Cray}} & \multicolumn{1}{c|}{\textbf{GNU}} & \multicolumn{1}{c|}{\textbf{LLVM}} & \textbf{NVIDIA} & \textbf{INTEL}   \\ \hline
\textbf{4.5}           & \multirow{5}{*}{C \& C++}   & \multicolumn{1}{c|}{146}          & \multicolumn{1}{c|}{142}           & \multicolumn{1}{c|}{137}          & 147           & \multicolumn{1}{c|}{142}           & \multicolumn{1}{c|}{145}          & \multicolumn{1}{c|}{148}           & 131             & 142              \\ \cline{1-1} \cline{3-11} 
\textbf{5.0}           &                          & \multicolumn{1}{c|}{179}          & \multicolumn{1}{c|}{147}           & \multicolumn{1}{c|}{170}          & 171           & \multicolumn{1}{c|}{146}           & \multicolumn{1}{c|}{175}          & \multicolumn{1}{c|}{172}           & 67              & 162              \\ \cline{1-1} \cline{3-11} 
\textbf{5.1}           &                          & \multicolumn{1}{c|}{68}           & \multicolumn{1}{c|}{39}            & \multicolumn{1}{c|}{75}           & 66            & \multicolumn{1}{c|}{39}            & \multicolumn{1}{c|}{75}           & \multicolumn{1}{c|}{66}            & 13              & 67               \\ \cline{1-1} \cline{3-11} 
\textbf{5.2}           &                          & \multicolumn{1}{c|}{16}           & \multicolumn{1}{c|}{1}             & \multicolumn{1}{c|}{12}           & 10            & \multicolumn{1}{c|}{1}             & \multicolumn{1}{c|}{18}           & \multicolumn{1}{c|}{10}            & 2               & 3                \\ \cline{1-1} \cline{3-11} 
\textbf{Total}         &                          & \multicolumn{1}{c|}{\textbf{409}} & \multicolumn{1}{c|}{\textbf{329}}  & \multicolumn{1}{c|}{\textbf{394}} & \textbf{394}  & \multicolumn{1}{c|}{\textbf{328}}  & \multicolumn{1}{c|}{\textbf{413}} & \multicolumn{1}{c|}{\textbf{396}}  & \textbf{213}    & \textbf{374}     \\ \hline
\textbf{4.5}           & \multirow{5}{*}{Fortran} & \multicolumn{1}{c|}{86}           & \multicolumn{1}{c|}{89}            & \multicolumn{1}{c|}{104}          & 15            & \multicolumn{1}{c|}{88}            & \multicolumn{1}{c|}{104}          & \multicolumn{1}{c|}{15}            & 97              & 97               \\ \cline{1-1} \cline{3-11} 
\textbf{5.0}           &                          & \multicolumn{1}{c|}{40}           & \multicolumn{1}{c|}{86}            & \multicolumn{1}{c|}{110}          & 9             & \multicolumn{1}{c|}{81}            & \multicolumn{1}{c|}{107}          & \multicolumn{1}{c|}{9}             & 24              & 85               \\ \cline{1-1} \cline{3-11} 
\textbf{5.1}           &                          & \multicolumn{1}{c|}{2}            & \multicolumn{1}{c|}{3}             & \multicolumn{1}{c|}{19}           & 0             & \multicolumn{1}{c|}{3}             & \multicolumn{1}{c|}{19}           & \multicolumn{1}{c|}{0}             & 2               & 12               \\ \cline{1-1} \cline{3-11} 
\textbf{5.2}           &                          & \multicolumn{1}{c|}{4}            & \multicolumn{1}{c|}{2}             & \multicolumn{1}{c|}{4}            & 3             & \multicolumn{1}{c|}{2}             & \multicolumn{1}{c|}{4}            & \multicolumn{1}{c|}{3}             & 3               & 3                \\ \cline{1-1} \cline{3-11} 
\textbf{Total}         &                          & \multicolumn{1}{c|}{\textbf{132}} & \multicolumn{1}{c|}{\textbf{180}}  & \multicolumn{1}{c|}{\textbf{237}} & \textbf{27}   & \multicolumn{1}{c|}{\textbf{174}}  & \multicolumn{1}{c|}{\textbf{234}} & \multicolumn{1}{c|}{\textbf{27}}   & \textbf{126}    & \textbf{197}     \\ \hline
\end{tabular}
\label{tab:perVersionResults}
\end{table}

\begin{figure}[h]
    \centering
    \includesvg[width=0.9\textwidth]{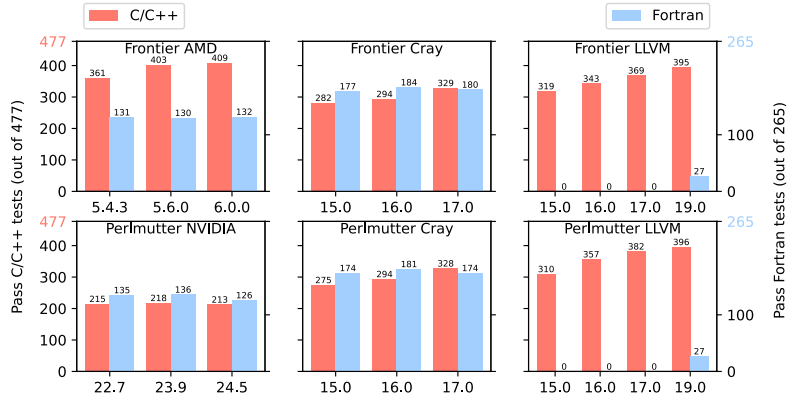}
    \caption{Evolution of compilers on Frontier and Perlmutter}
    \label{fig:evolution-pass-tests}
\end{figure}

Figure \ref{fig:evolution-pass-tests} shows the compiler evolution over a period of two years for Perlmutter and Frontier\footnote{\url{zenodo.org/doi/10.5281/zenodo.12571032}: compilerVersionsPerlmutterFrontierResults.json}. The chosen compilers are the recommended compilers in the given system for OpenMP offloading application development. It can be seen in the graph that the compiler development for Fortran has been slow during this period. In the case of C/C++ there has been greater improvement over time, specially for LLVM. NVIDIA supports till OpenMP 4.5. 

\subsection{SPEChpc HPG Benchmarking Suite}
\label{sec:spec}
We present SPEChpc V1.1.8 HPG~\cite{SPEChpc} benchmark for evaluating the performance and quality of OpenMP offloading features~\cite{9826013}. The benchmark contains 9 MPI applications written in different programming models, including OpenMP-offloading (TGT) and OpenACC (ACC), where 6 of them are C/C++ and 3 are Fortran.
While we focus on the OpenMP offloading versions of the applications, we also present the OpenACC results as a performance reference of another directive programming model.
Table \ref{tab:specApps} summarizes the SPEChpc applications. We use the smallest size (tiny) versions of the benchmarks, which can run on a single node of each of the tested systems. In addition to these tests, we include a modified \emph{532.sph\_exa}, which we will reference as \emph{532.sph\_exaM}. The modified version was a result of code analysis as we noticed a bigger performance difference with the OpenACC version compared to OpenACC (we discuss this in more detail at the latter half of this section).
\begin{table}[!ht]
    \caption{SPEChpc 2021 version 1.1.8 applications taken from SPEC website~\cite{SPEChpc}.} 
\centering
\begin{tabular}{|l|c|c|l|}
\hline
\multicolumn{1}{|c|}{\textbf{Benchmark}}     & \multicolumn{1}{c|}{\multirow{2}{*}{\textbf{Size: Tiny}}}   & \multicolumn{1}{c|}{\multirow{2}{*}{\textbf{Language}}} & \multicolumn{1}{c|}{\multirow{2}{*}{\textbf{Application Area}}}                 \\ 
\multicolumn{1}{|c|}{\textbf{Name}}                       &    &    &                  \\ \hline \hline
\textbf{LBM D2Q37}                            & 505.lbm      & C                 & Computational Fluid Dynamics              \\ \hline
\textbf{SOMA}                                 & 513.soma     & C                 & Polymeric Systems               \\ \hline
\textbf{Tealeaf}                              & 518.tealeaf  & C                 & High Energy Physics             \\ \hline
\textbf{Cloverleaf}                           & 519.clvleaf  & Fortran           & High Energy Physics             \\ \hline
\textbf{Minisweep}                            & 521.miniswp  & C                 & Nuclear Engineering \\ \hline
\textbf{POT3D}                                & 528.pot3d    & Fortran           & Solar Physics                             \\ \hline
\textbf{SPH-EXA}                              & 532.sph\_exa & C++14             & Astrophysics and Cosmology                \\ \hline
\textbf{HPGMG-FV}                             & 534.hpgmgfv  & C                 & Cosmology, Astrophysics     \\ \hline
\textbf{miniWeather}                          & 535.weather  & Fortran           & Weather                                   \\ \hline
\end{tabular}
    \label{tab:specApps}
\end{table}

\textbf{SPEChpc Results:} Table \ref{tab:specResults} presents estimated base time results of SPEChpc 2021 Version 1.1.8 on Frontier\footnote{\url{zenodo.org/doi/10.
5281/zenodo.12571032}: Frontier\_SPEC\_PaperResults.zip} and Perlmutter\footnotemark{} 
using available compilers. 

\begin{table}[!ht]
    \caption{Estimated {\bf Base Run Time} of SPEChpc tiny applications on Perlmutter and Frontier using one node. \textbf{EE} is execution error, while \textbf{BE} is build error.} 
    \centering
\begin{tabular}{|l|c|c|c|c||c|c|c|c|}
\hline
    & \multicolumn{8}{c|}{\textbf{Estimate Base Time (seconds)}}  \\ \cline{2-9} 
    & \multicolumn{4}{c||}{\textbf{Perlmutter}}  & \multicolumn{4}{c|}{\textbf{Frontier}}    \\ \hline
\textbf{Compiler}           & \textbf{GNU}  & \textbf{LLVM}   & \textbf{Cray}   & \textbf{NVIDIA} & \textbf{GNU}  & \textbf{LLVM}   & \textbf{Cray}   & \textbf{ROCm}  \\ 
\textbf{Version}     & \textbf{14.1} & \textbf{19.0.0} & \textbf{17.0.0} & \textbf{24.5}   & \textbf{14.1} & \textbf{19.0.0} & \textbf{17.0.0} & \textbf{6.0.0} \\ \hline \hline 
\textbf{505.lbm}         & 484.89        & 38.29           & \textit{28.34}           & 35.9            & 2813.46       & 43.44           & 40.82           & 54.64          \\ \hline
\textbf{513.soma}        & 855.05        & 69.61           & \textit{56.75}           & 65.64           & \textbf{BE}          & 87.98           & \textbf{BE}            & 70.05          \\ \hline
\textbf{518.tealeaf}     & 2200.95       & 90.84           & 49.09           & 40.49           & 337.12        & 43.58           & \textit{40.41}           & 48.51          \\ \hline
\textbf{519.clvleaf}     & \textbf{BE}          & \textbf{BE}            & \textbf{EE}            & \textit{45.54}           & \textbf{BE}          & \textbf{BE}            & 58.73           & 72.73          \\ \hline
\textbf{521.miniswp}     & \textbf{EE}          & 209.55          & 96.76           & 573.09          & \textbf{EE}          & 160.44          & \textit{93.59}           & 142.61         \\ \hline
\textbf{528.pot3d}       & 926.24        & \textbf{BE}            & 55.34           & 61.54           & \textbf{BE}          & \textbf{BE}            & \textit{45.64}           & 92.61          \\ \hline
\textbf{532.sph\_exa}    & 1454.46       & 849.6           & \textbf{EE}            & 491.41          & \textbf{BE}          & \textit{203.34}          & 226.33          & 207.4          \\ \hline
\textbf{532.sph\_exaM} & 5973.46       & 179.41          & \textit{128.36}          & \textbf{EE}            & \textbf{BE}          & 145.61          & 164.87          & 144.83         \\ \hline
\textbf{534.hpgmgfv}     & \textbf{EE}          & 156.75          & \textit{71.2}            & 163.33          & \textbf{BE}          & 102.32          & 87.5            & 95.59          \\ \hline
\textbf{535.weather}     & 1391.84       & \textbf{BE}            & 38.51           & 42.72           & 2569.96       & \textbf{BE}            & \textit{32.51}           & 53.19          \\ \hline
\end{tabular}
    \label{tab:specResults}
   \vspace{-4mm}
\end{table}

Looking at the results from the perspective of the applications, only \textbf{505.lbm} and \textbf{518.tealeaf} can be built and executed with all compilers (both are C/C++ applications). Fortran's \textbf{519.clvleaf} can only be built and executed by NVIDIA's compiler in Perlmutter and by AMD's ROCm and Cray compilers in Frontier. 
From the point of view of the compilers, only AMD's ROCm compiler could build and execute all applications. The performance of LLVM C/C++ applications is similar to that from vendor compilers. LLVM Flang is under development so it could not build Fortran applications. Even though the GNU compiler can build Fortran and C/C++ applications, its performance is very low compared to all others, specially on Frontier, where many of the applications don't even build. This is an interesting observation, as we found GNU compilers doing a better job with correctness (Figure~\ref{fig:pass-tests}).

As mentioned before, AMD's compiler was the only one that built and executed all applications with good performance. For this to happen, AMD required special flags suggested by the developers, like (\emph{-fopenmp-target-xteam-reduction-blocksize=128}, \emph{-Mx,201,2}, \emph{-fno-openmp-assume-no-nested-parallelism}, \emph{--mca topo basic}). Finally, as a performance reference for the Perlmutter system, we present SPEChpc results for the OpenACC programming model\footnotemark[\value{footnote}] as shown in Table \ref{tab:specAccResults}. We noticed significant performance differences between OpenMP and OpenACC outputs for some applications, especially \textbf{532.sph\_exa} and \textbf{521.miniswp}.
As mentioned before, a direct comparison of results from both the models should not and cannot be made without understanding the intricate details of feature implementations and understanding the reasoning behind performance gaps. 
\footnotetext{\url{zenodo.org/doi/10.5281/zenodo.12571032}: Perlmutter\_SPEC\_PaperResults.zip}

Listing \ref{lst:sphexaacc} contains the OpenMP annotation added to \textbf{532.sph\_exa}. We found that this memory allocation on the device was missing for the OpenMP offloading version, while it was present for the OpenACC version. The modified version is referenced in this document as \textbf{532.sph\_exaM}\footnote{\url{zenodo.org/doi/10.
5281/zenodo.12571032}: SqPatch.hpp.diff}. In the original OpenMP version, because these pragmas were missing, every time these arrays were accessed in a target region they have to be allocated in the device. By adding these lines, the arrays are allocated in the device before multiple iterations of device kernel calls. In Perlmutter, while the original version does not execute correctly with Cray compiler (it is a memory allocation error), the modified version doesn't run correctly with NVIDIA and GNU Compilers (it is also a memory allocation error). In the case of Frontier, neither versions compile with GNU, but the error doesn't have anything to do with the modification, it is a compiler bug.

\begin{lstlisting}[frame=tlrb, language=C, caption={These are the pragmas added to the benchmark \textbf{532.sph\_exa}. The new benchmark is \textbf{532.sph\_exaM}. These lines were added to the original file \textbf{SqPatch.hpp} on the function \textbf{resizeN(size\_t size)}.}, label=lst:sphexaacc]
// OpenACC annotation present in the original code
#pragma acc exit data delete(hNptr, hNCptr)
// First OpenMP Directive added
#pragma omp target exit data map(delete: hNptr[:Nsze], 
                                         hNCptr[:NCsze])
                                 
// OpenACC annotation present in the original code
#pragma acc enter data create(hNptr[:size*ngmax],
                              hNCptr[:size])
// Second OpenMP Directive added
#pragma omp target enter data map(alloc: hNptr[:size*ngmax], 
                                         hNCptr[:size])

                                         
\end{lstlisting}


\begin{table}[!ht]
    \centering
\caption{OpenACC estimated {\bf base run time} of SPEChpc tiny applications on Perlmutter with NVIDIA compiler.} 
\begin{tabular}{|l|ccccccccc|}
\hline
              & \multicolumn{9}{c|}{\textbf{Estimate Base Time (seconds)}} \\ \hline
\multicolumn{1}{|c|}{\multirow{2}{*}{\textbf{Version}}} & \multicolumn{1}{c|}{\textbf{505}} & \multicolumn{1}{c|}{\textbf{513}} & \multicolumn{1}{c|}{\textbf{518}} & \multicolumn{1}{c|}{\textbf{519}} & \multicolumn{1}{c|}{\textbf{521}} & \multicolumn{1}{c|}{\textbf{528}} & \multicolumn{1}{c|}{\textbf{532}} & \multicolumn{1}{c|}{\textbf{534}} & \textbf{535} \\ 
   & \multicolumn{1}{c|}{\textbf{lbm}} & \multicolumn{1}{c|}{\textbf{soma}} & \multicolumn{1}{c|}{\textbf{tealeaf}} & \multicolumn{1}{c|}{\textbf{clvleaf}} & \multicolumn{1}{c|}{\textbf{miniswp}} & \multicolumn{1}{c|}{\textbf{pot3d}} & \multicolumn{1}{c|}{\textbf{sph\_exa}} & \multicolumn{1}{c|}{\textbf{hpgmgfv}} & \textbf{weather} \\ \hline
\textbf{24.5}       & \multicolumn{1}{c|}{28.48}               & \multicolumn{1}{c|}{45.82}                & \multicolumn{1}{c|}{48.23}                   & \multicolumn{1}{c|}{35.69}                   & \multicolumn{1}{c|}{52.38}                   & \multicolumn{1}{c|}{53.58}                 & \multicolumn{1}{c|}{129.08}                   & \multicolumn{1}{c|}{64.27}                   & 37.23                   \\ \hline
\end{tabular}
    \label{tab:specAccResults}
\end{table}

\textbf{Discussion:} We observed that it should not be concluded that one compiler is better over the other as the results largely depends on the applications and optimizations. Having said that, Cray demonstrated some of the best results. 
Furthermore, Cray's results on Perlmutter are very similar to the OpenACC results for \textbf{505.lbm}, \textbf{528.pot3d} and \textbf{535.weather}. Moreover, the results of \textbf{532.sph\_exaM} for Cray compiler is similar to the OpenACC version of \textbf{532.sph\_exa}. These also suggest that the large performance gaps for codes such as  \textbf{521.miniswp} between OpenMP and OpenACC could be an implementation difference, not necessarily a compiler optimization difference.
SPEChpc uses only 4.5 specification features, but as seen, the compilers are still having trouble, after almost 10 years of the release of the specification, to produce efficient code for all applications. Moreover, the performance is not as good as the OpenACC versions. As seen with \textbf{532.sph\_exa}, the lack of performance of the applications may not be caused only by compiler optimization issues, but it could also be application implementation differences as a result of not using OpenMP features that compilers don't support or that their performance is low. 
The performance of larger versions of SPEChpc would correlate to the results shown here, but they would also depend on the network, and the performance of the MPI implementation. We expect that the performance per node would be similar, but we can not extrapolate to the performance of larger versions of the benchmark.

\section{Conclusion}
\label{sec:conclusion}
This paper sheds light on the pass/fail of OpenMP features from version 4.5 and above. We do so by building a CI/CD workflow comprising of OpenMP V\&V, Smoke, HeCBench, and SPEChpc suites to automate the testing of OpenMP implementations with a goal to fail fast and provide feedback to LLVM compiler developers especially since several vendor compilers are LLVM-based. This paper further provides details of the current status of correctness of OpenMP compilers via the V\&V testing infrastructure and tests running on Frontier, Perlmutter, and Sunspot.  As we know, manual test creation is a time consuming process for developers; this challenge is currently being explored by using Large Language Models (LLM) for test generation~\cite{MUNLEY20241}. 
The work also discusses the quality of OpenMP compiler implementations on different GPUs when using SPEChpc benchmarking suite. The discussion also offers suggestions and room for further implementation improvement.

Building on the workflow developed in this work, future works could include the implementation of "expected fail" testing, the continous development of tests as the specification evolves, and enabling developers to reproduce and debug problems effectively.

In conclusion, the OpenMP offloading features represent a crucial environment for the HPC community as the systems continue to evolve. The model has been bringing legacy and new applications to run on novel systems, which makes it time critical to closely track correctness and quality of implementations guiding the developers accordingly. 



\textbf{Acknowledgments}
This research used resources of the OLCF at ORNL supported by the Office of Science of the U.S. DOE under Contract No. DE-AC05-00OR22725; used resources of the ALCF, a U.S. DOE Office of Science user facility at Argonne National Laboratory and is based on research supported by the U.S. DOE Office of Science-ASCR program, under Contract No. DE-AC02-06CH11357; used resources of NERSC, a U.S. DOE Office of Science User Facility located at LBNL, operated under Contract No. DE-AC02-05CH11231 using NERSC ERCAP0029463. This material is also based upon work supported by the U.S. DOE under Contract DE-FOA-0003177, S4PST: Next Generation Science Software Technologies Project. 

%
%
%
\bibliographystyle{splncs04}
\bibliography{references}
%
\end{document}